\documentclass{ws-mpla}

\usepackage[super]{cite}
\usepackage{xcolor}
\usepackage[verbose,hypertexnames=false]{hyperref}
\definecolor{blue}{rgb}{0.0, 0.0, 1.0}
\definecolor{red}{rgb}{1.0, 0.0, 0.0}
\definecolor{royalblue}{rgb}{0.0, 0.14, 0.4}
\hypersetup{colorlinks=true,linkcolor=red, citecolor=blue, urlcolor=blue}
\def\orcid#1{\kern .08em\href{https://orcid.org/#1}{\includegraphics[keepaspectratio,width=0.7em]{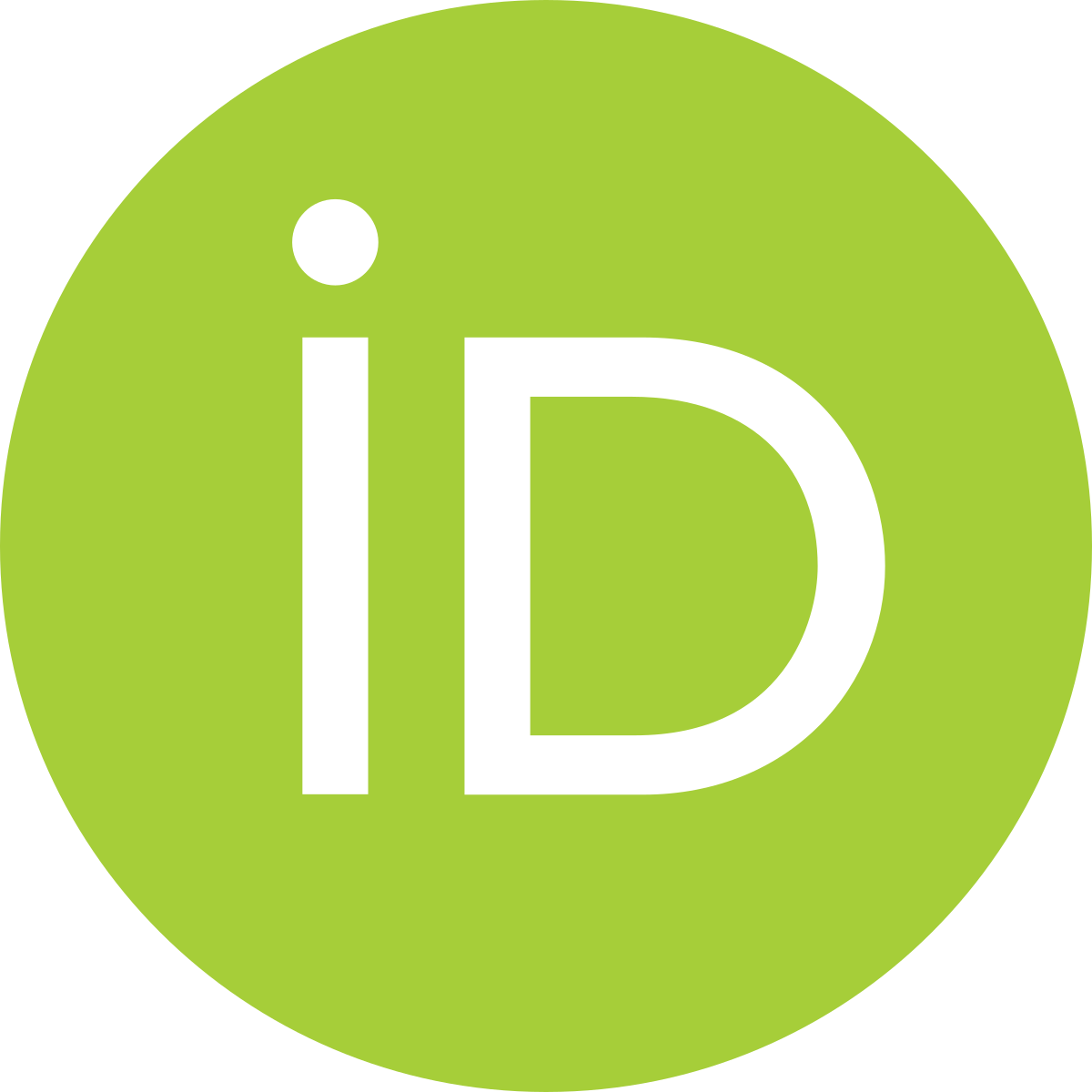}}}
\begin{document}

\markboth{S. Puhan, N. Kumar, H. Dahiya}{Internal structure of light mesons using the power law wave function}

%%%%%%%%%%%%%%%%%%%%% Publisher's Area please ignore %%%%%%%%%%%%%%
\catchline{}{}{}{}{}
%%%%%%%%%%%%%%%%%%%%%%%%%%%%%%%%%%%%%%%%%%%%%%%%%%%%%%%%%%%%%%%%%%%
%----------------------------------------------------------------
\title{Internal structure of light mesons using the power law wave function}
%----------------------------------------------------------------

%----------------------------------------------------------------
\author{Satyajit Puhan\orcid{0009-0003-8583-4054}}
\address{Computational High Energy Physics, Department of Physics, Dr. B R Ambedkar National Institute of Technology, Jalandhar, 144008, India \\
E-mail Address: puhansatyajit@gmail.com}
\author{Narinder Kumar\orcid{0000-0002-5481-1162}}
\address{Computational Theoretical High Energy Physics Lab,\\
Department of Physics,\\
Doaba College, Jalandhar 144004, India\\
E-Mail Address: narinderhep@gmail.com}
\author{Harleen Dahiya\orcid{0000-0002-3288-2250}}
\address{Computational High Energy Physics, Department of Physics, Dr. B R Ambedkar National Institute of Technology, Jalandhar, 144008,India \\
E-mail Address:dahiyah@nitj.ac.in}
%----------------------------------------------------------------
\maketitle

%\begin{history}
%\received{(Day Month Year)}
%\revised{(Day Month Year)}
%\accepted{(Day Month Year)}
%\published{(Day Month Year)}
%\end{history}

%----------------------------------------------------------------
\begin{abstract}
In this paper, we study the internal structure of light pseudoscalar mesons using spin-improved power-law wave functions. We chose the pion and the kaon for our work. We use the standard quark-quark correlation functions to calculate the distribution amplitudes (DAs), parton distribution functions (PDFs), transverse momentum parton distribution functions (TMDs), and generalized parton distribution functions (GPDs) at zero skewness and form factors. We present all the above distribution functions through the overlap of light-front wave functions (LFWFs).

We use leading-order Efermov-Radyushkin-Brodksy-Lepage (ERBL) equations for DAs and next-to-leading-order (NLO) Dokshitzer–Gribov–Lipatov–Altarelli–Parisi (DGLAP) equations for PDFs to evolve them to higher scales. We find that only $41\%$ of the longitudinal momentum fraction is carried by the quark and antiquark of both pion and kaon at $16$ GeV$^2$. The vector form factors for both the pion and the kaon are found to be in good agreement with the experimental data. Similarly, the electromagnetic charge radii are found to be $0.668$ fm and $0.704$ fm for pion and kaon, respectively.
\end{abstract}
%----------------------------------------------------------------

\keywords{Power law wave functions, Distribution amplitudes, parton distribution functions, form factors, Transverse momentum parton distribution functions.}
%\ccode{PACS Nos.: 03.65.$-$w, 04.62.+v}

%-----------------------------------------------------------------
\section{Introduction}	
%-----------------------------------------------------------------
It is always a challenging task to understand the internal structure of hadrons from the first principles of quantum chromodynamics (QCD) \cite{Pich:1995ua}. The internal constituents, quarks, antiquarks, gluons, and sea-quarks play an important role in describing the perturbative and non-perturbative properties of the hadrons. There have been a lot of theoretical and experimental studies for the distribution of these constituents. The study of light mesons like the pion and the kaon plays an important role in describing the spontaneously broken chiral symmetry in QCD. Much work is being done to describe the distribution of the internal structure of the pion and kaon through theoretical models, lattice simulations, and experiments. The internal structure can be studied using the multi-dimensional generalized transverse momentum parton distribution functions (GTMDs), generalized parton distribution functions (GPDs), transverse momentum parton distribution functions (TMDs) and parton distribution functions (PDFs) using the quark-quark correlation matrix \cite{Puhan:2025kzz,deTeramond:2018ecg,Meissner:2008ay,Aguilar:2019teb}.

In this study, we have used a non-perturbative power law wave function (PLWF) to study the different distribution functions for the ground state of pion and kaon by considering the meson as a bound state of a quark-antiquark pair. This method was first proposed by Schlumpf \cite{Schlumpf:1994bc} and has been used to study the hadron structure in Refs. \cite{Choi:1997qh,Melikhov:1995xz}. We have studied the leading twist distribution amplitudes (DAs), TMDs, PDFs, GPDs and form factors (FFs) for the case of pion and kaon. The choice of the power law wave function is largely phenomenological. However, its use is motivated by several phenomenological and theoretical considerations. It is well known fact that, Gaussian wave function falls off rapidlly at large ${\bf k_\perp}$ which makes it inconsistent with the expected asymptotic momentum distribution of quarks inside hadrons. In contrast, PLWF naturally reproduced the correct large momentum tail and for the same reason it also provides the better description of hadronic form factors at large $Q^2$. \cite{Lepage:1980fj, Efremov:1979qk, Chernyak:1983ej,Brodsky:1973kr}  \\
For the case of pseudoscalar mesons, there is only one longitudinal DA present at the leading twist compared to two DAs for the vector meson case. Similarly, a single unpolarized quark TMD and GPD is present at the leading twist for the pseudoscalar mesons, which provides the three-dimensional structure of the quark inside the hadron. The only quark PDF of the pseudoscalar mesons can be obtained from the TMD and GPD with proper relations. While the charge form factor has been derived from the unpolarized GPD by integrating over the longitudinal momentum fraction. As we are calculating all these structure functions in the non-perturbative region, so we have evolved them to the higher scale to compare with the available experimental results. The leading twist DA of pion and kaon have been evolved using the leading order (LO) ERBL evolutions \cite{Puhan:2024xdq}, while the PDF have been evolved using the next-to-leading order (NLO) DGLAP evolutions \cite{Miyama:1995bd}. These results obtained in this work are going to be studied in future experiments such as the Electron-Ion Collider (EIC) at BNL~\cite{Arrington:2021biu}, the Electron-ion collider in China (EicC)~\cite{Anderle:2021wcy}, J-PARC~\cite{Sawada:2016mao}, the upgraded JLAB 12 GeV~\cite{Accardi:2023chb}, and the COMPASS/AMBER++ at CERN~\cite{Adams:2018pwt}.\\
This paper is organized as follows. We first present the bound state meson wave function of quark antiquark in section~\ref{sec:NJL}. In this section, we have also discussed the power law wave function along with the spin wave function and the input parameters used in this work. section~\ref{sec:das} presents the calculations and results of the pion and kaon DAs along with their decay constants. In section \ref{sec:gpd}, we have calculated and discussed the results of pion and kaon unpolarized GPDs. We have also calculated the respective charge form factors and radii for both mesons in this section. We have demonstrated the pion and kaon leading twist TMDs and PDFs in section~\ref{sec:tmds}. Finally, we summarize in section \ref{sec:sum}.
%----------------------------------------------------------------
\section{Power Law Wave Functions} \label{sec:NJL}
%----------------------------------------------------------------
Light-front dynamics provides a suitable framework for describing the hadron structure in terms of its constituents. The light-cone (LC) Fock-state of a hadron corresponding to a quark-antiquark pair can be expressed as \cite{Puhan:2023ekt,Belyaev:1997iu}
\begin{eqnarray}
|\pi(K)(P^+,P_\perp)\rangle &=& \sum_{\lambda_1,\lambda_2}\int
\frac{\mathrm{d} x \mathrm{d}^2
        \mathbf{k}_{\perp}}{\sqrt{x(1-x)}16\pi^3}
           \Psi_{\pi(K)}(x,\mathbf{k}_{\perp},\lambda_1,\lambda_2)|x,\mathbf{k}_{\perp},
        \lambda_1,\lambda_2 \rangle
        .
        \label{meson}
\end{eqnarray}
Here, $|\pi(K)\rangle$ is the hadron eigenstate of pion (kaon) with $P=(P^+,P_\perp)$ being the four vector momenta, $k=(k^+,\textbf{k}_\perp)$ is the four vector momenta of quark, $x=k^+/P^+$ is the longitudinal momentum fraction carried by the quark from its parent hadron, $\lambda_{1(2)}$ is the helicity of the quark (antiquark) and $\Psi_{\pi(K)}(x,\mathbf{k}_{\perp},\lambda_1,\lambda_2)$ is the total wave function of the meson which can be expressed as 
\begin{eqnarray}
    \Psi_{\pi(K)}(x,\mathbf{k}_{\perp},\lambda_1,\lambda_2)= \phi(x,\mathbf{k}_{\perp}) \times \chi(x,\mathbf{k}_{\perp},\lambda_1,\lambda_2).
    \label{spin}
\end{eqnarray}
Here, $\phi(x,\mathbf{k}_{\perp})$ is the momentum space wave function of the hadron, and for this work, we have considered the PLWFs as \cite{Choi:1997qh}
\begin{eqnarray}
\phi(x,\mathbf{k}_{\perp})=N_{\pi(K)}\Bigg(1+\frac{((x-1/2)M_{\pi(K)}+\frac{m^2_{\bar q}-m_q^2}{2 M_{\pi(K)}})^2+\mathbf{k}^2_{\perp}}{\beta^2_{\pi(K)}}\Bigg)^{-s}.
\label{plwf}
\end{eqnarray}
Here, the power $s$ is typically taken to be $2$. $M_{\pi(K)}$ is the bound state mass of the pion (kaon) with $m_{q(\bar q)}$ being the quark (antiquark) mass. $N$ and $\beta$ correspond to the normalization constant and the harmonic scale parameter, respectively. In this work, we have considered $m_q=0.2$ GeV and $\beta_\pi=0.41$ GeV for the pion. For kaon, the quark masses are taken to be $m_q=0.22$ and $m_{\bar q}=0.45$ along with $\beta=0.405$ \cite{Puhan:2023ekt}. $\chi(x,\mathbf{k}_{\perp},\lambda_1,\lambda_2)$ in Eq. (\ref{spin}) is the spin wave function and can be expressed in the form of a quark-meson vertex as \cite{Puhan:2024jaw}
\begin{eqnarray}
\chi(x,\mathbf{k}_{\perp},\lambda_1,\lambda_2)=\bar u (k_1,\lambda_1) \frac{\gamma_5}{\sqrt{2}\sqrt{{M_{\pi(K)}}^2-(m^2_q-m^2_{\bar q}})} v(k_2,\lambda_2).
\end{eqnarray}
Here, $u$ and $v$ are the Dirac spinors with momentum $k_1$ and $k_2$, respectively.
%----------------------------------------------------------------
\section{Distribution Amplitudes} \label{sec:das}
%----------------------------------------------------------------
At large momentum transfer, DAs can be analyzed through the exclusive processes. One can easily access the light-cone distributions via light-front wave functions by integrating out the transverse momentum. The correlation for defining the pseudoscalar DAs can be expressed as follows \cite{Li:2017mlw,Choi:2007yu}
\begin{equation}
   \langle 0|\bar{\xi}(z) \gamma^+ \gamma_5 \xi(-z)| \pi(K)(P^+,{\bf P_\perp}) \rangle = i k^+ \kappa_{\pi(K)}\int_{0}^{1} dx \, e^{i(x-1/2) k^+ z^- } \phi_{\pi(K)} (x) \bigg|_{z^+,\textbf{z}_\perp=0} \, ,
\label{DA} 
\end{equation} 
where $z=(z^+,z^-,{\bf z_\perp})$ is the position four vector. $ \xi$ represents the quark field operator and $\kappa_\pi$ is the decay constant. In substituting the meson state from Eq. (\ref{meson}) and quark field operators in Eq. (\ref{DA}), one can have the DAs in the form of LFWFs as 
\begin{equation}
    \frac{\kappa_{\pi(K)}}{2 \sqrt{2 N_c}} \phi_{\pi(K)} (x)=\frac{1}{\sqrt{2x(1-x)}} \int \frac{d^2 \textbf{k}_\perp}{16 \pi^3} [\Psi_{\pi(K)}(x,\textbf{k}_\perp,\uparrow,\downarrow)-\Psi_{\pi(K)}(x,\textbf{k}_\perp,\downarrow,\uparrow)] \, ,
\end{equation}
with $N_c=3$ being the number of colors of a quark flavor. $\kappa_{\pi(K)}$ is the decay constant of pion (kaon). The decay constant of the pion is found to be $114$ MeV compared to the experimental results of $130.2\pm 1.7 $ MeV \cite{ParticleDataGroup:2022pth}, whereas for the case of the kaon, the decay constant is found to be $144$ GeV (PDG value is $156.1 \pm 0.5$ MeV). We found that the decay constants of pion and kaon are found to have an error of $12.44\%$ and $7.69\%$ from the PDG data, respectively. We have also observed that the choice of $s$ value in Eq. (\ref{plwf}) is very sensitive to the decay constant.
The pion DA is normalized as 
\begin{eqnarray}
    \int_{0}^{1} dx \, \phi_{\pi(K)}(x)=1 \, .
\end{eqnarray}
\begin{figure}[ht]
\centering
\includegraphics[width=0.485\linewidth]{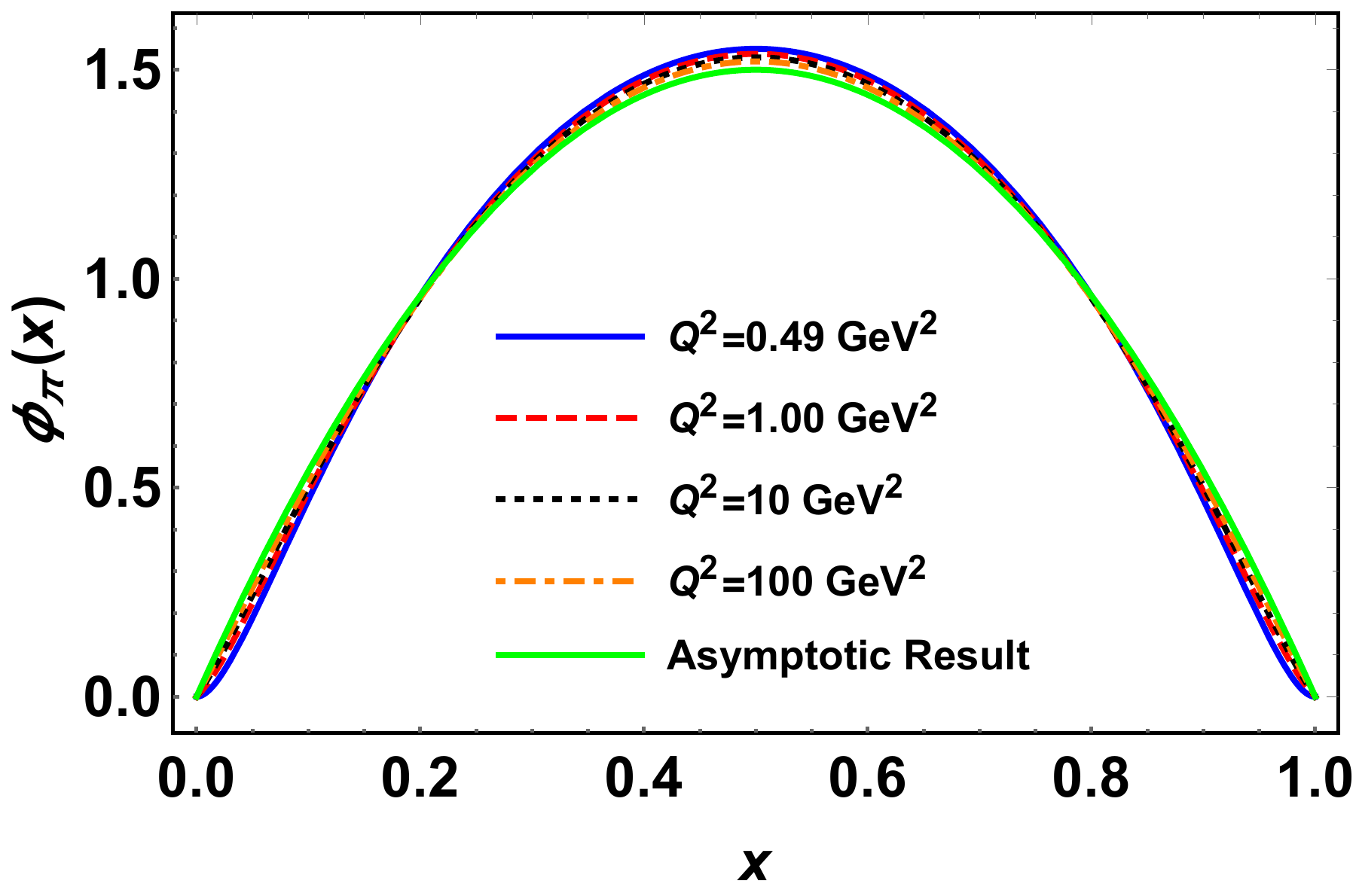}
\includegraphics[width=0.485\linewidth]{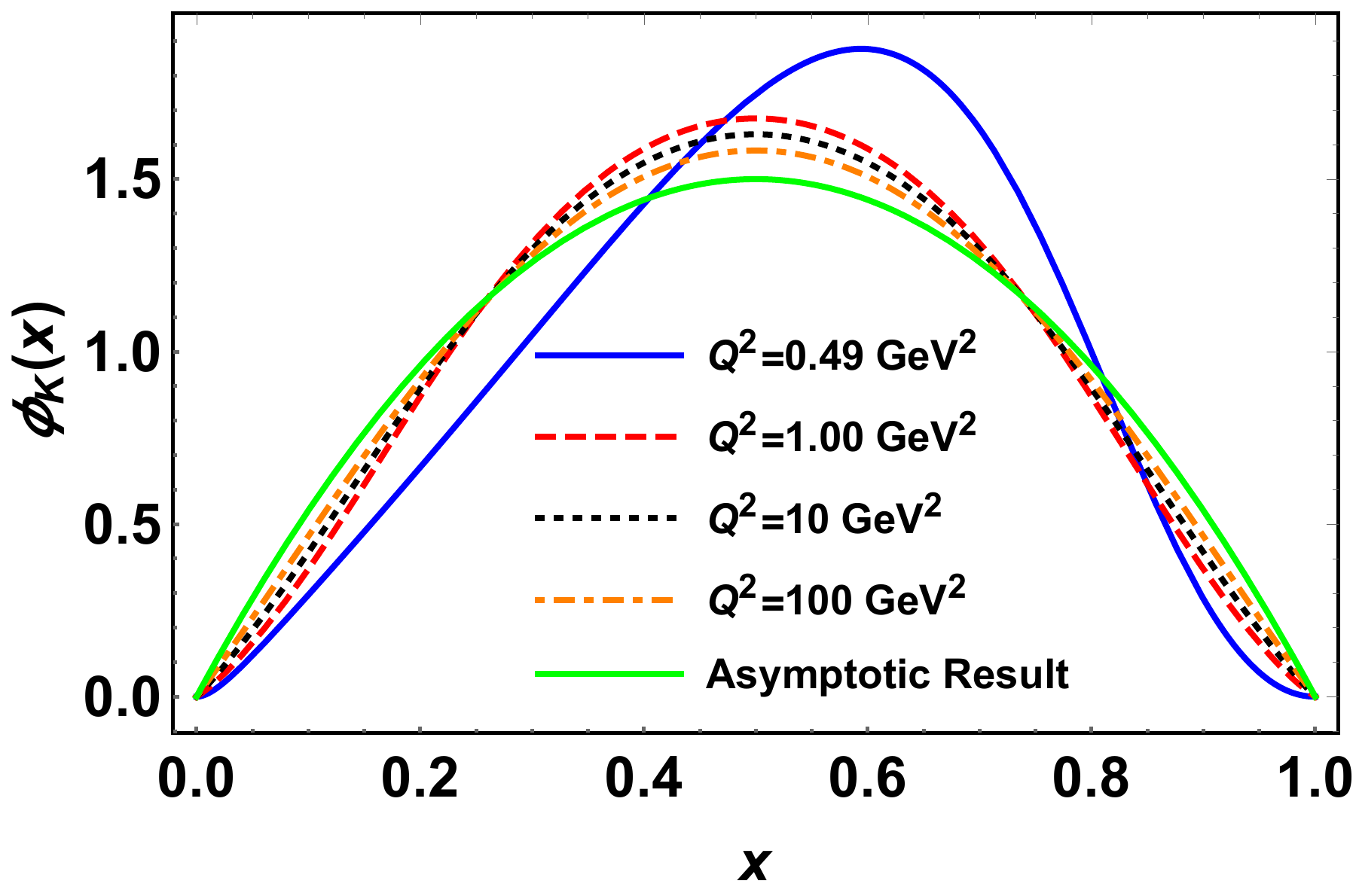}
\caption{The DAs of pion and kaon plotted with respect $x$ at model scale as well as at $Q^2=1$, $10$ and $100$ GeV$^2$ along with comparison with asymptotic freedom $6x (1-x)$ in the left and right panels, respectively.}
\label{p1}
\end{figure}
The DAs of pion and kaon have been plotted in Fig. \ref{p1} at an initial scale of $Q_0^2=0.49$ GeV$^2$ along with different higher scales using the LO ERBL evolutions. The pion DAs are found to be symmetric at each scale and in sync with the asymptotic results of $6 x(1-x)$. For the case of kaon, the DA shows a shift towards the high $x$ values.
\begin{figure}[ht]
\centering
\includegraphics[width=0.485\linewidth]{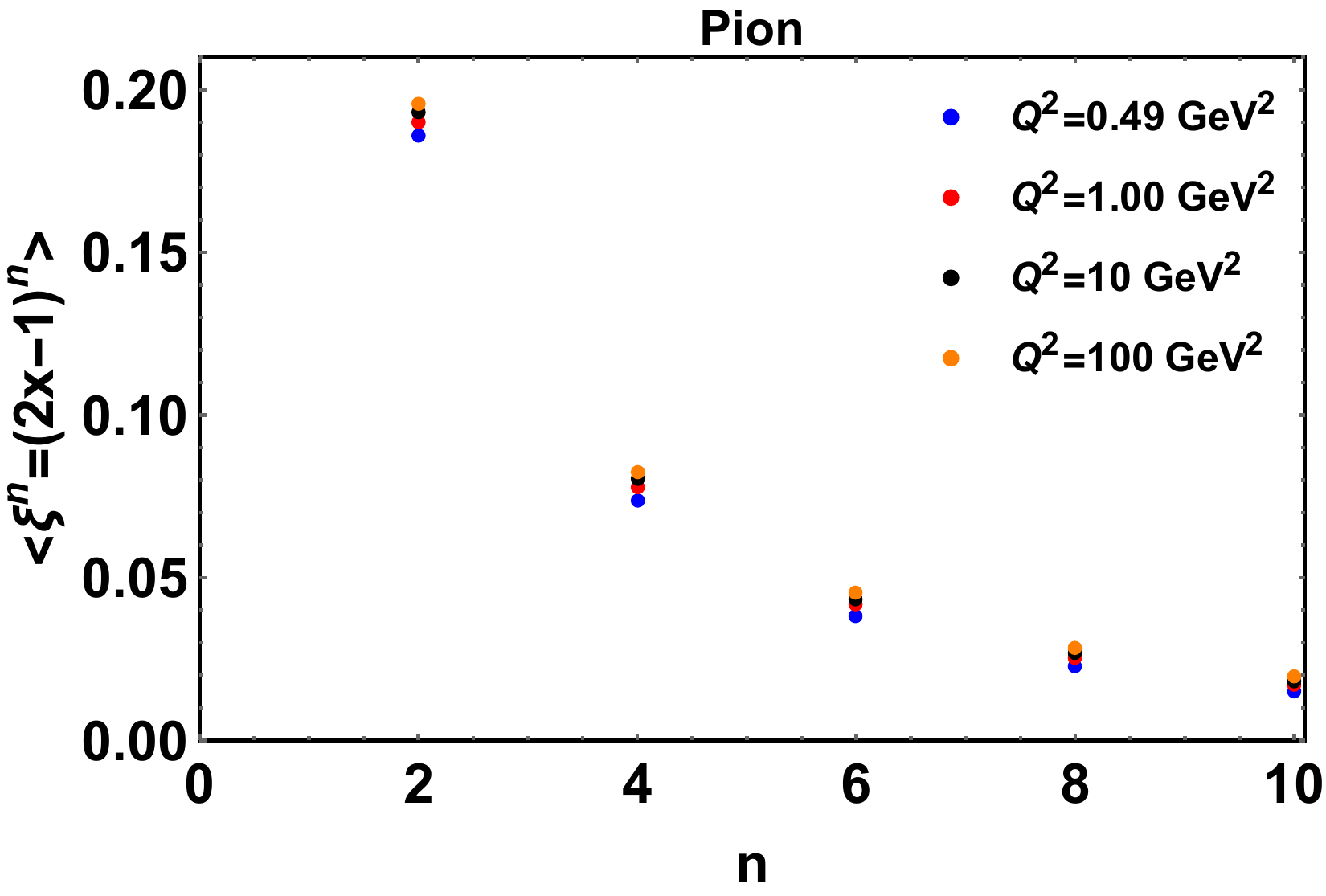}
\includegraphics[width=0.485\linewidth]{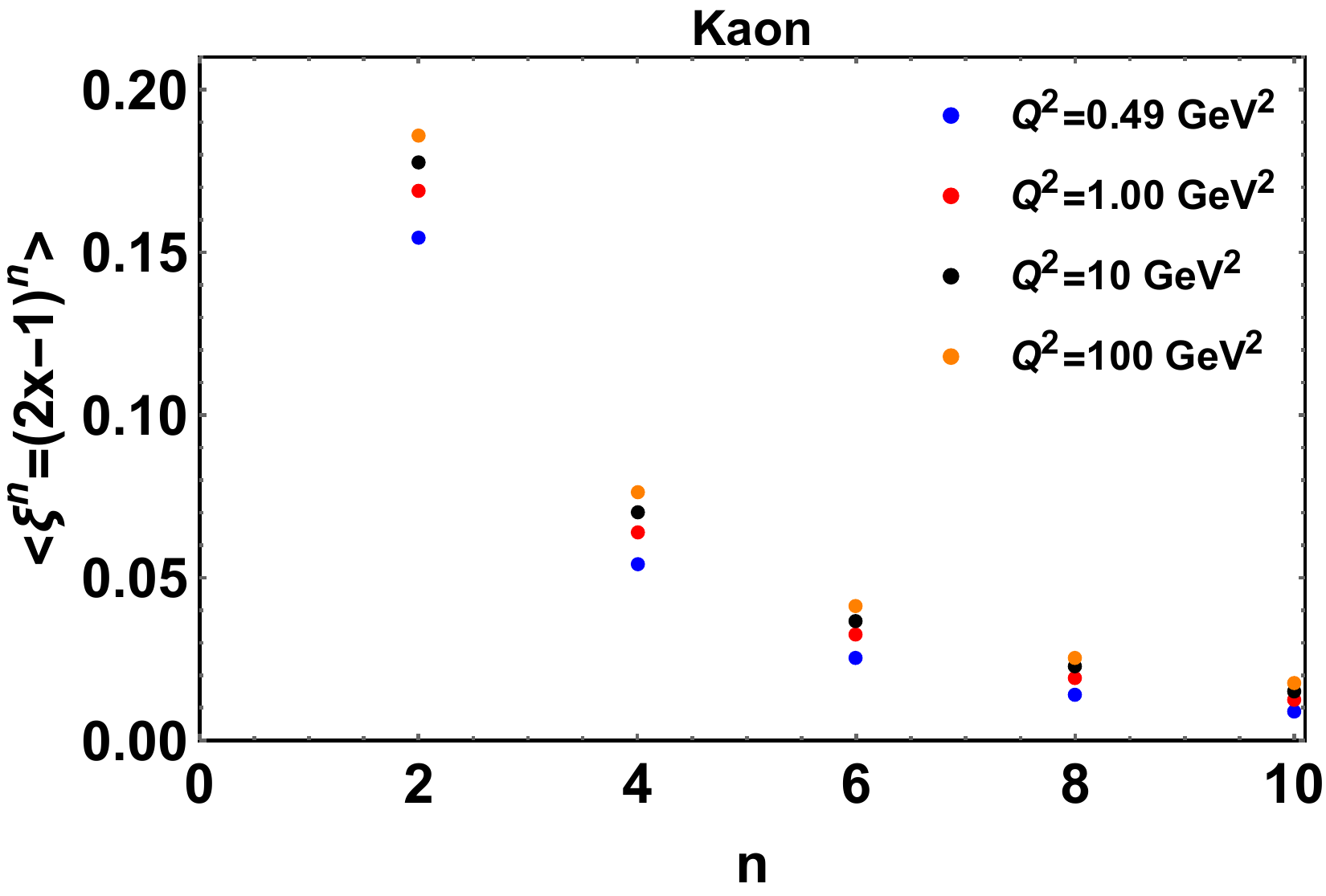}
\caption{The Mellin moment $<\xi^n>$ of the DAs plotted up to n=10 for pion and kaon in the left and right panels, respectively. }
\label{p2}
\end{figure}
We have also presented the results in Fig. \ref{p2} for the Mellin moment $\langle \xi^n \rangle$ of pion and kaon DAs. The Mellin moment $\langle \xi^n \rangle$ of the DAs can be calculated as 
\begin{eqnarray}
    \langle \xi^n \rangle =\int_0^1 dx \ (1-2x)^n \phi_{\pi(K)}(x).
\end{eqnarray}
%----------------------------------------------------------------
\section{Generalized Parton Distributions} \label{sec:gpd}
%----------------------------------------------------------------
The matrix elements of the quark operators at a light-like separation are defined as GPDs \cite{Diehl:2003ny}. For spin-$0$ particles, we have only one chiral-even unpolarized GPD, which can be defined in terms of the bilocal current as
\begin{eqnarray}
   H_{\pi(K)}^q(x,\xi,t)&=&\frac{1}{2} \int \frac{dz^-}{2\pi} e^{ix\bar{P} \cdot z} \nonumber\\
  &&\times \bigg\langle \pi(K)(P^{+\prime \prime},{\bf P^{\prime \prime}_\perp})\bigg|~\bar{\xi} \big(-\frac{z}{2}\big)~ \gamma^+ \xi \big(\frac{z}{2}\big) ~\bigg| \pi(K)(P^{+\prime},{\bf P^{\prime}_\perp}) \bigg\rangle_{z^+= {\bf z_\perp}=0} \, . 
    \label{tmdeq}
\end{eqnarray}
\begin{figure}[ht]
\centering
\includegraphics[width=0.485\linewidth]{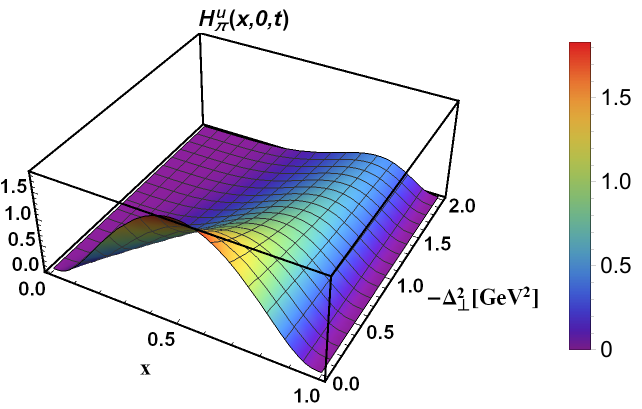}
\includegraphics[width=0.485\linewidth]{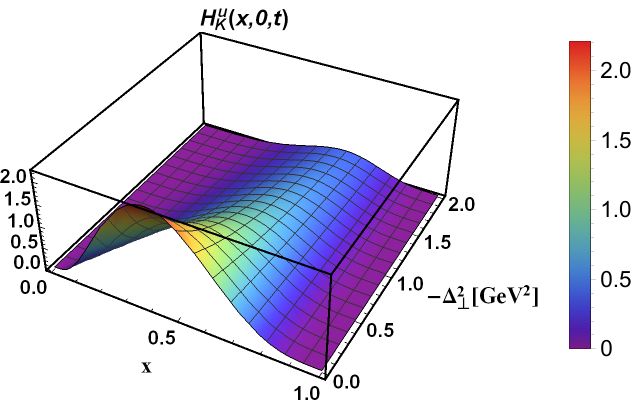}
\caption{The unpolarized quark GPD of pion and kaon plotted with respect to $x$ and $\Delta_\perp^2$ GeV$^2$ at zero skewness for pion and kaon in the left and right panel, respectively.}
\label{p3}
\end{figure}
Other kinematic variables, which include the four-momentum transfer and skewness parameter, are respectively expressed as $\Delta^\mu=P^{\prime \prime \mu}-P^{\prime \mu}$ with $t=\Delta^2=-\Delta_\perp$ and $\zeta=-\Delta^+/2P^+$. We have chosen the light-front gauge as $A^+=0$, which in turn makes the gauge link, appearing between the quark field operators, unity. The overlap form of GPD $H^q_{\pi(K)}(x,0,t)$ with zero skewness can be expressed as
\begin{eqnarray}
    H^q_{\pi (K)}(x,0,t)&=&\int \frac{d^2 \mathbf{k_\perp}}{16 \pi^3} \big[\Psi_{\pi(K)}^\ast (x^{\prime\prime},\mathbf{k}^{\prime\prime}_\perp, \uparrow, \uparrow) \Psi_{\pi(K)} (x^{\prime},\mathbf{k}^{\prime}_\perp, \uparrow, \uparrow) \nonumber \\ 
    &+& \Psi_{\pi(K)}^\ast (x^{\prime\prime},\mathbf{k}^{\prime\prime}_\perp, \uparrow, \downarrow) \Psi_{\pi(K)} (x^{\prime},\mathbf{k}^{\prime}_\perp, \uparrow, \downarrow) \nonumber\\
    &+& \Psi_{\pi(K)}^\ast (x^{\prime\prime},\mathbf{k}^{\prime\prime}_\perp, \downarrow, \uparrow) \Psi_{\pi(K)} (x^{\prime},\mathbf{k}^{\prime}_\perp, \downarrow, \uparrow) \nonumber\\ &+&  \Psi_{\pi(K)}^\ast (x^{\prime\prime},\mathbf{k}^{\prime\prime}_\perp, \downarrow, \downarrow)
    \Psi_{\pi(K)} (x^{\prime},\mathbf{k}^{\prime}_\perp, \downarrow, \downarrow)\big] \, ,
\label{GPDeq}
\end{eqnarray}
where $\textbf{k}_\perp^{\prime\prime}$ and $\textbf{k}_\perp^{\prime}$ correspond to the final and initial state quark momentum, respectively. In a symmetric frame, they can be expressed as
\begin{eqnarray}
    \textbf{k}_\perp^{\prime\prime}&=&\textbf{k}_\perp-(1-x^{\prime\prime})~\frac{\Delta_\perp}{2} \, , \nonumber \\
    \textbf{k}_\perp^{\prime}&=&\textbf{k}_\perp+(1-x^{\prime})~\frac{\Delta_\perp}{2} \, .
\end{eqnarray}
Since we are dealing with the zero skewness GPDs, the initial and final state longitudinal momentum fraction carried by an active quark of a meson remains the same. In Fig. \ref{p3}, we have presented the results for the pion and kaon GPDs. We observe that the pion GPD remains symmetric over $x$ as both quarks carry the same mass. Further, broad distribution in $t=-\Delta_\perp^2$ indicates the compact transverse structure. In the case of the kaon GPD, where one quark is lighter and the other heavier, the distribution is asymmetric, and the peak is shifted towards small values of $x$. A comparison between these two plots indicates that differences in $t$ dependence lead to modifications of the transverse localization due to a heavier quark mass. \\
\begin{figure}[ht]
\centering
\includegraphics[width=0.485\linewidth]{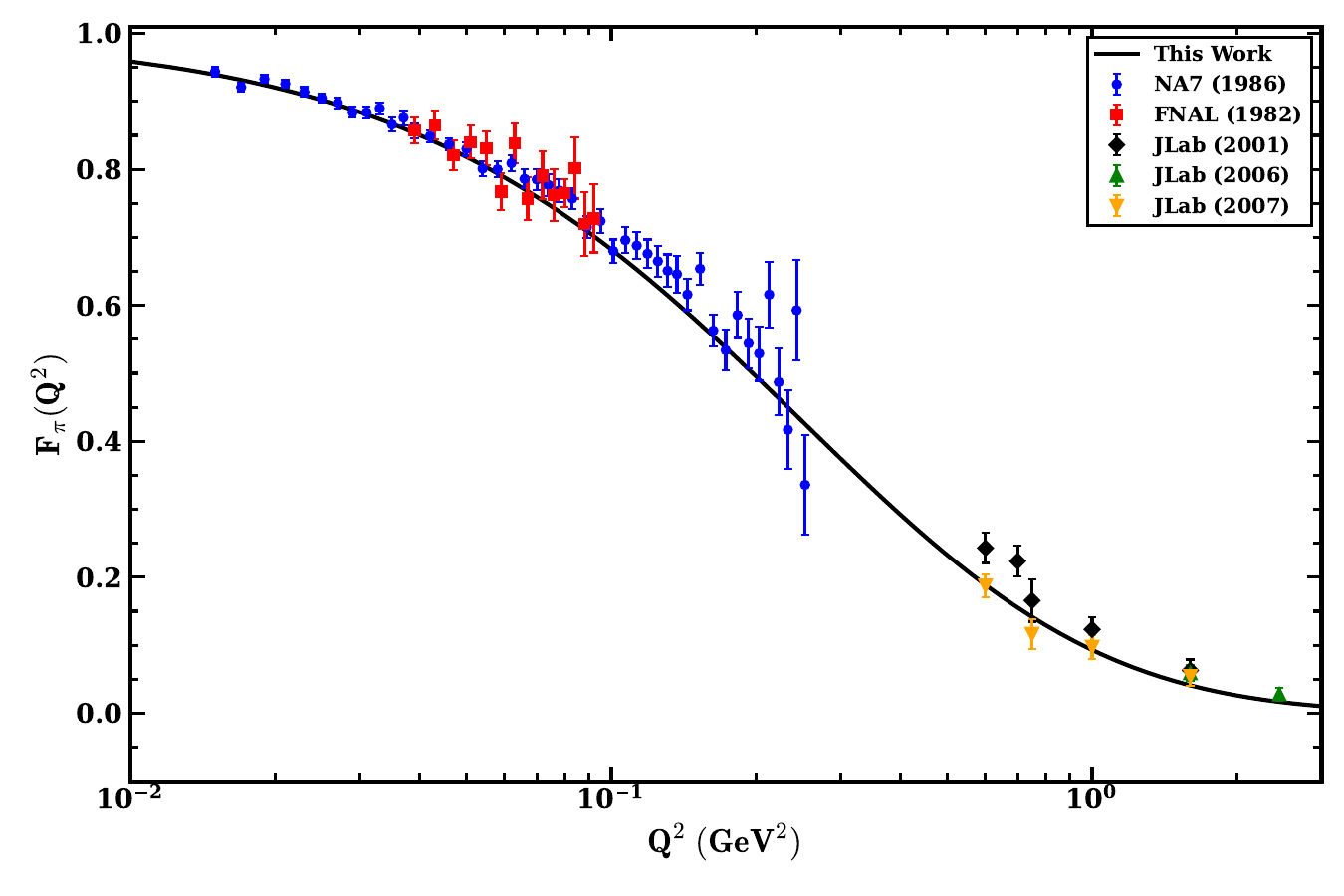}
\includegraphics[width=0.485\linewidth]{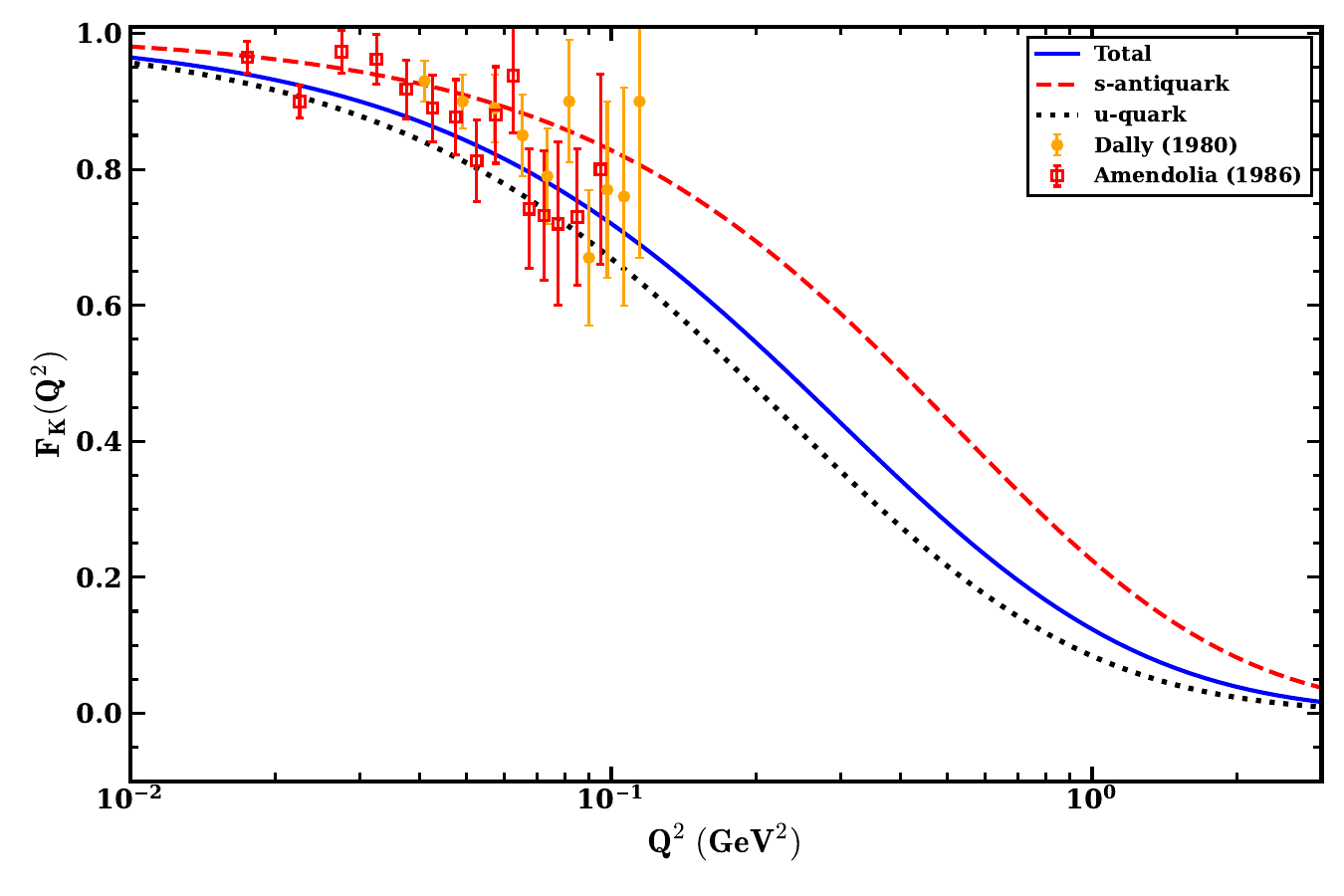}
\caption{The charge FFs of pion and kaon (along with constituent quark antiquark) plotted with respect to $Q^2$ in the left and right panels, respectively. We have also compared our results with the available experimental results. \cite{NA7:1986vav,Dally:1982zk,JeffersonLabFpi:2000nlc,JeffersonLabFpi-2:2006ysh,JeffersonLabFpi:2007vir,Dally:1980dj,Amendolia:1986ui}}
\label{ffs}
\end{figure}
The charge form factor $F^{\pi(K)}(Q^2)$ that corresponds to the unpolarized GPD can also be calculated as \cite{Puhan:2025pfs} 
\begin{eqnarray}
    F^{\pi(K)}(Q^2)= e_q \int_0^1 dx H^q_{\pi (K)}(x,0,t) + e_{\bar q}\int_0^1 dx H^{\bar q}_{\pi (K)}(x,0,t).\label{ffq}
\end{eqnarray}
Here, $e_{q(\bar q)}$ is the quark (antiquark) charge.
In Fig. \ref{ffs}, we have presented the results for the charge form factors of pion and kaon, respectively. For the pion case, we have compared our results with the experimental data from JLab, NA7, and FNAL collaborations \cite{NA7:1986vav,Dally:1982zk,JeffersonLabFpi:2000nlc,JeffersonLabFpi-2:2006ysh,JeffersonLabFpi:2007vir}. We observed that our results are in very good agreement at both low and high $Q^2$ values. Similarly, in the right-hand panel, we compare our results for contributions towards kaon EMFF coming from the strange antiquark and the up quark. We have also compared our results with the experimental data whose data points are available for low $Q^2$ \cite{Dally:1980dj,Amendolia:1986ui}. Our result is consistent with the data points. For the kaon case, we infer that due to the presence of a heavier strange quark charge distribution is more compact compared to the pion case. \\
The charge radii of the pion and the kaon can be calculated from the charge form factors as
\begin{eqnarray}
    \langle r_{\pi(K)}^2\rangle &= & \frac{-6}{F_{\pi(K)}(0)} \frac{\partial F_{\pi(K)}(Q^2)}{\partial Q^2}\Big|_{Q^2\rightarrow0}.
\end{eqnarray}
The charge radii of pion and kaon are found to be $0.668$ and $0.704$ fm, which are very close to the experimental values of $0.657$ and $0.583$ fm \cite{NA7:1986vav,Amendolia:1986ui} respectively. The errors of the radius of the pion and kaon are found to be $1.7 \%$ and $20.7\%$, respectively, from the experimental values.
%----------------------------------------------------------------
\section{Transverse Momentum Dependent Parton Distributions} \label{sec:tmds}
%----------------------------------------------------------------
At leading-twist, there are two quark TMDs which are present for pseudoscalar mesons, out of which one is T-even in nature \cite{Meissner:2008ay}. The T-even quark TMDs can be calculated using the quark-quark correlator. The quark-quark correlation function leading-twist TMDs for spin-$0$ hadrons can be expressed as \cite{Mulders:1995dh}
\begin{eqnarray}
    \Phi^{(\Gamma)}(x,\mathbf{k}_{\perp}^2)&=&\frac{1}{2} \int \frac{dz^- d^2z_\perp}{2 (2\pi)^3} e^{ix\bar{P} \cdot z} \nonumber\\
    &\times& \bigg\langle \pi(K)(P^+,P_\perp)\bigg|~\bar{\xi} \big(\frac{-z}{2}\big)~\mathcal{W}(\frac{-z}{2},\frac{z}{2}) \Gamma \xi \big(\frac{z}{2}\big) ~\bigg| \pi(K)(P^+,P_\perp) \bigg\rangle \, .
    \label{tmdeq}
\end{eqnarray}
Here, $\Gamma=\gamma^+$ corresponds to $f_1(x,\textbf{k}^2_\perp)$ and  $\mathcal{W}(-z/2,z/2)$ is the Wilson line (taken to be unity for this case). The leading-twist unpolarized TMD can be obtained from the above correlator as
\begin{eqnarray}
    \Phi^{(\gamma^+)}(x,\mathbf{k}_{\perp}^2) =f_1(x,\textbf{k}^2_\perp).
\end{eqnarray}
\begin{figure}[ht]
\centering
\includegraphics[width=0.485\linewidth]{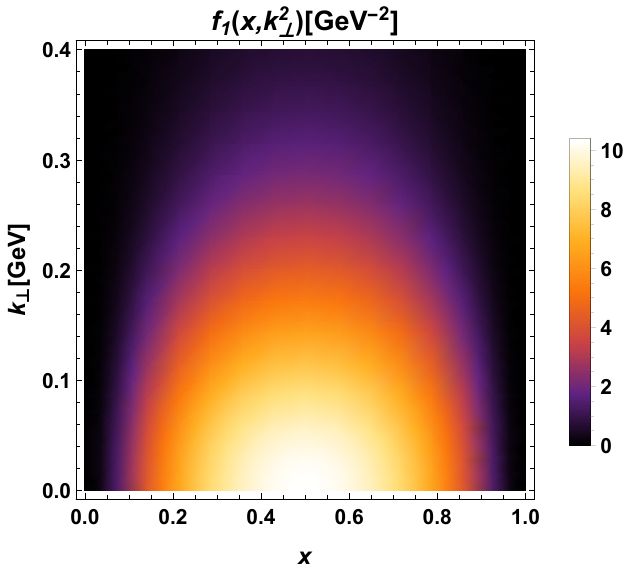}
\includegraphics[width=0.485\linewidth]{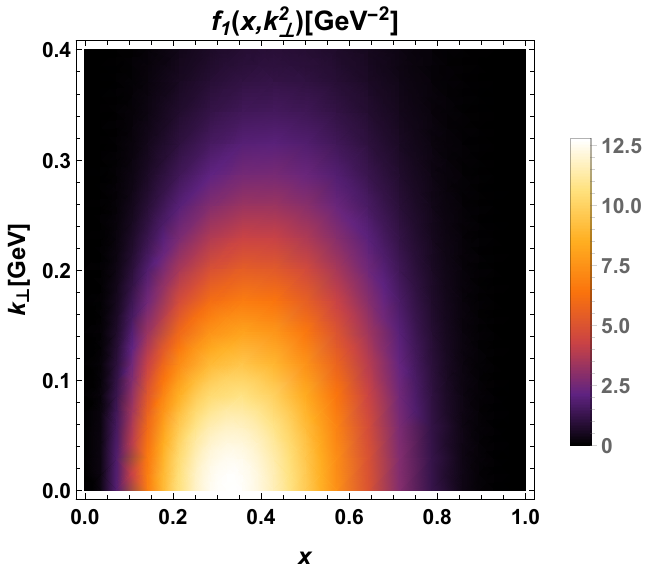}
\caption{The leading twist unpolarized quark TMDs plotted with respect to $x$ and $\textbf{k}_\perp$ GeV for pion and kaon in the left and right panels, respectively.}
\label{unpoltmd}
\end{figure}
By solving the correlator of Eq. (\ref{tmdeq}) using the pion and kaon wave functions of Eq. (\ref{meson}), the overlap form of unpolarized TMD is found to be 
\begin{eqnarray}
    f_1(x,\textbf{k}^2_\perp)&=&\frac{1}{16\pi^3} \sum_{\lambda_1,\lambda_2}\Bigg[\Psi^{*}_{\pi(K)}\left(x, \boldsymbol{k}_{\perp},\lambda_1,\uparrow\right) \Psi_{\pi(K)}\left(x, \boldsymbol{k}_{\perp},\lambda_2,\uparrow\right)\nonumber\\
    && + \Psi^{*}_{\pi(K)}\left(x, \boldsymbol{k}_{\perp},\lambda_1,\downarrow\right) \Psi_{\pi(K)}\left(x, \boldsymbol{k}_{\perp},\lambda_2,\downarrow\right) \Bigg].
\end{eqnarray}
In Fig. \ref{unpoltmd}, we have presented the density plots for unpolarized TMD $f_1$ for both pion and kaon, respectively. Some show common features are observed in the plots, which include a peak at a low value of $k_\perp$ and suppression at large $k_\perp$. This clearly indicates that both mesons carry a soft and non-perturbative bound state. The vanishing of density at the endpoints of $x$ implies that the probability is suppressed.  Further, in the case of the pion TMD, the symmetric distribution and the peak near $x \approx 0.5$ reflect that both quarks share almost the same amount of momentum. However, for kaon TMD, the peak is shifted towards $x \approx 0.3$ where the heavier quark carries  a large amount of momentum fraction, and the light quark peaks at a small value of $x$. This asymmetry in mass distribution distorts the momentum sharing relative to the pion TMD case. 
\begin{figure}[ht]
\centering
\includegraphics[width=0.485\linewidth]{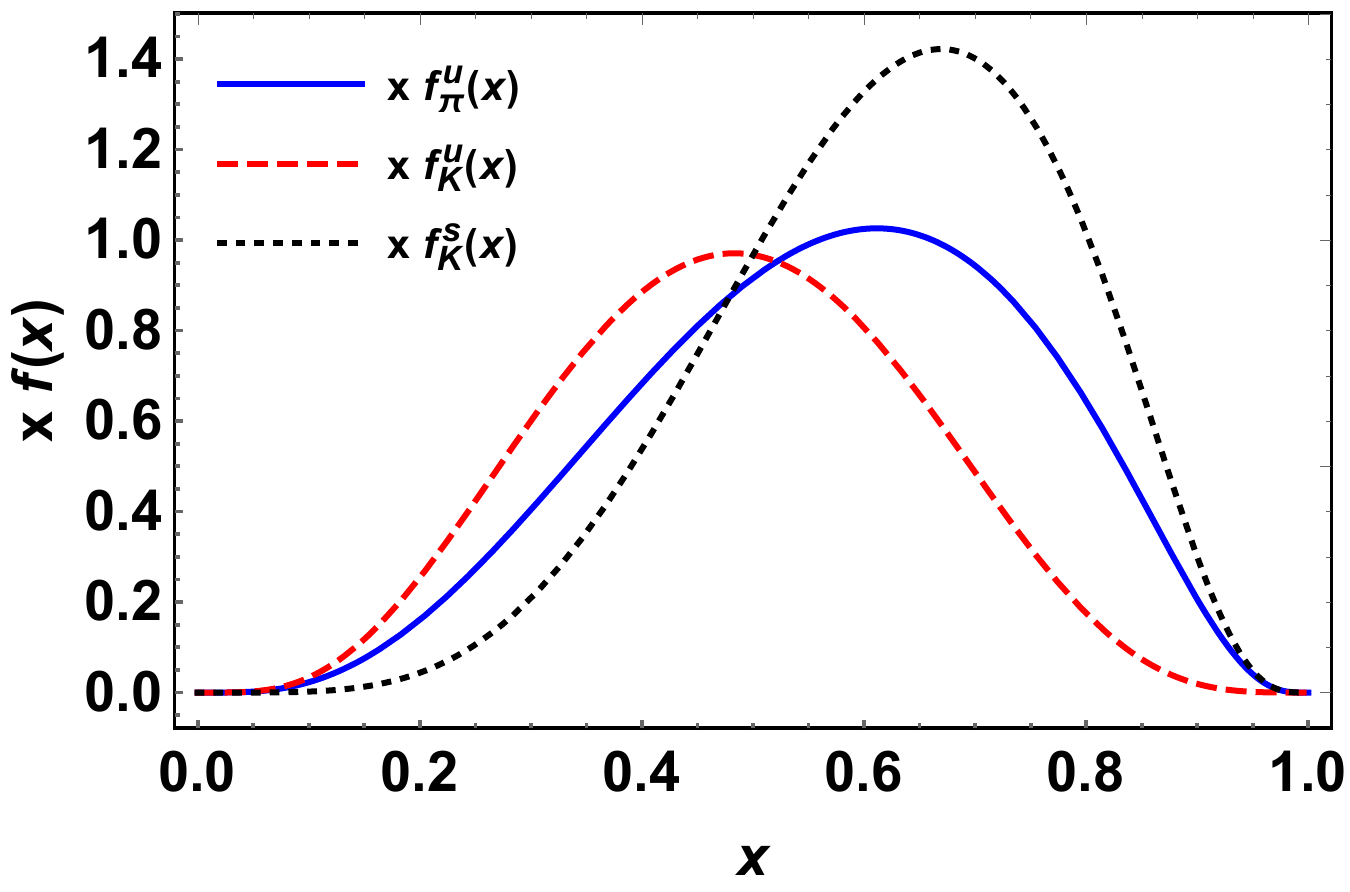}
\includegraphics[width=0.485\linewidth]{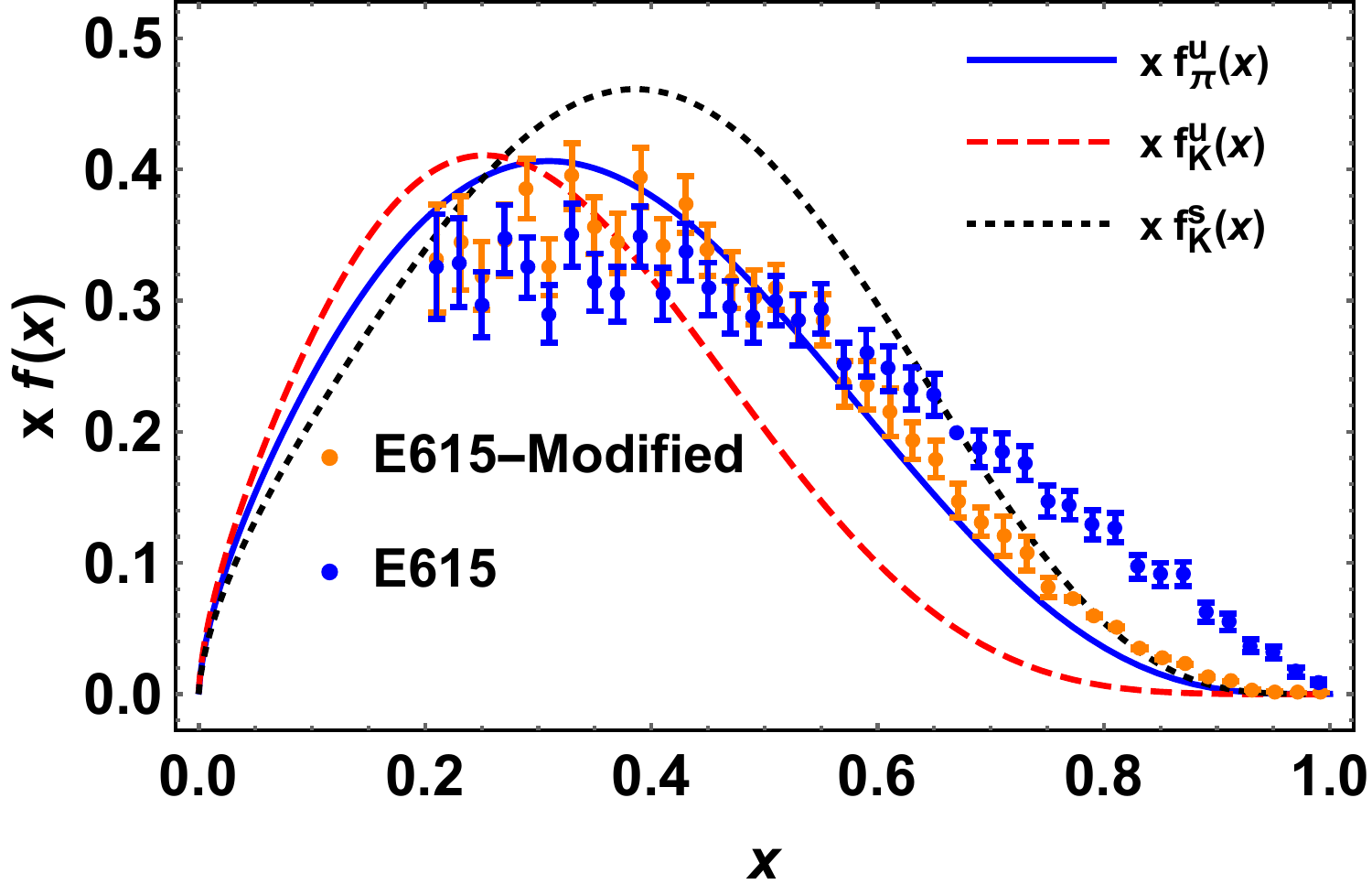}
\caption{The unpolarized quark PDFs of pion and kaon plotted at the model scale $Q_0^2=0.2$ GeV$^2$ and at 16 GeV$^2$ in the left and right panels, respectively. We have also compared our pion u-quark PDF results with available E-615 results \cite{Aicher:2010cb,E615:1989bda}.}
\label{pdfs}
\end{figure}
For the case of spin-0 hadrons, there is only one unpolarized quark PDF at the leading twist, which can be calculated from the GPD and TMD as
\begin{eqnarray}
    f(x)=\int_0^{\infty} d^2 \textbf{k}_\perp f_1(x,\textbf{k}_\perp^2)=H^q_{\pi(K)}(x,0,0).
\end{eqnarray}
In Fig. \ref{pdfs}, we have presented results for the unevolved (left panel) and DGLAP evolved (right panel) PDFs of the pion and kaon (for both u quark and s antiquark). In the left panel, as discussed previously, it is clear that for the pion, the momentum distribution is symmetric, but for the case of the kaon, because the strange quark is much heavier than the up quark, it hogs a large amount of momentum. For the up quark, peaks shift towards low values of $x$. In the right panel, we have evolved the PDFs to a higher energy scale $Q^2= 16$ GeV$^2$ through NLO DGLAP evolution using numerical code on a brute-force method \cite{Miyama:1995bd}, so that we can compare our results with experimental data from the E615 experiment at Fermilab \cite{Aicher:2010cb,E615:1989bda}. We observe that peaks are shifted towards lower values of $x$, and the distribution is broadened. This is due to the fact that, as the energy scale increases, the quark radiates gluons by losing its momentum and hence softens the distribution. Our results for the pion are in good agreement with the E615-modified data.
\begin{figure}[ht]
\centering
\includegraphics[width=0.485\linewidth]{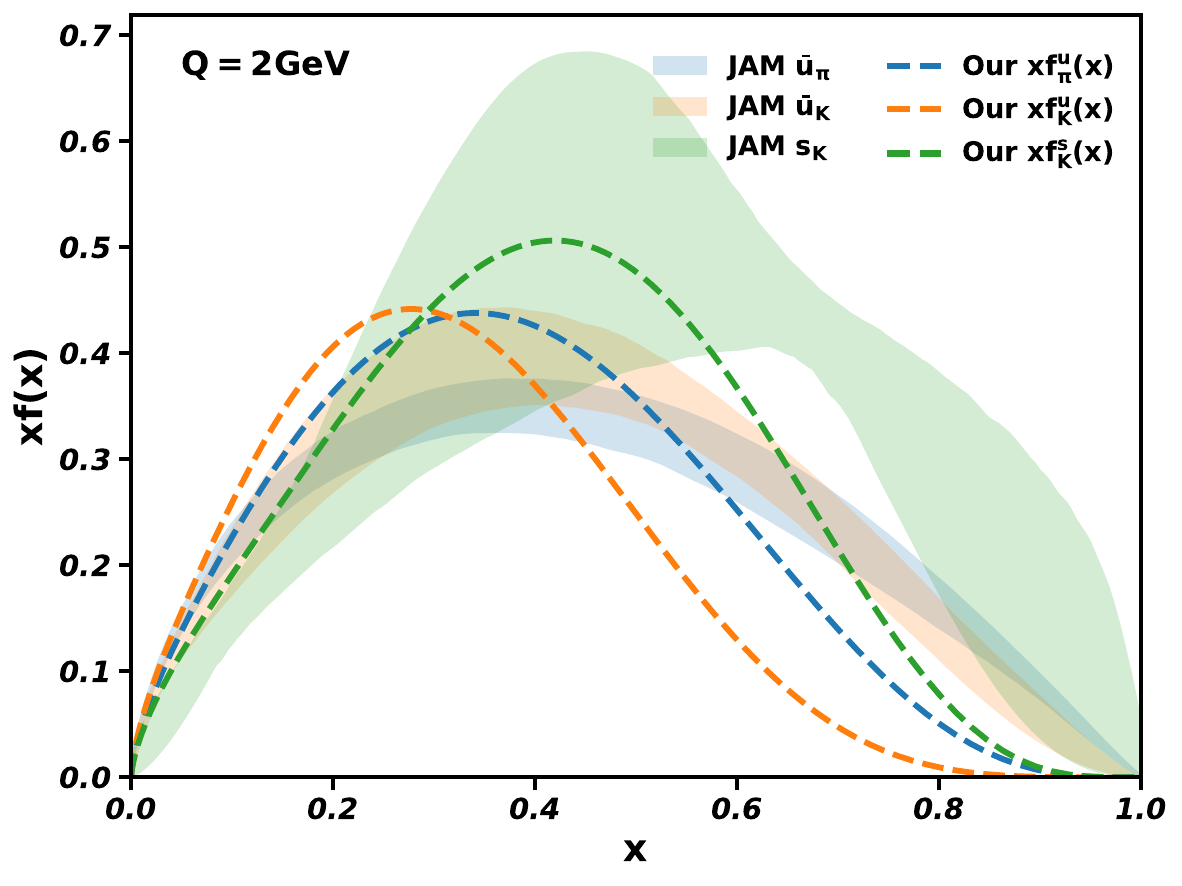}
\includegraphics[width=0.485\linewidth]{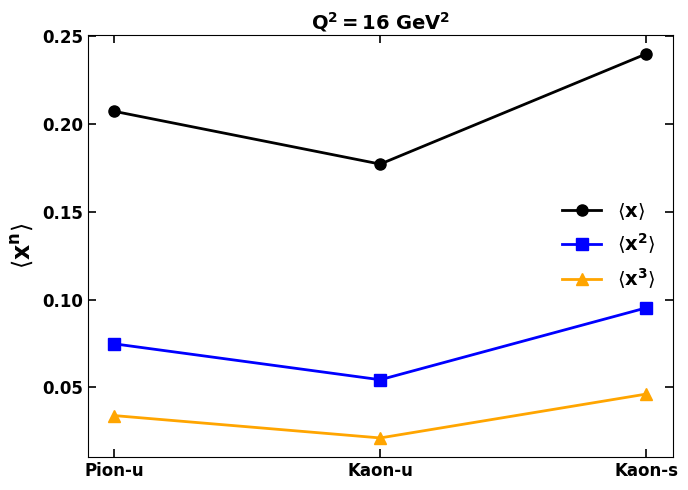}
\caption{(Left panel) The unpolarized quark PDFs of pion and kaon plotted at the model scale $Q=2$ GeV with comparison with recent JAM data \cite{Barry:2025wjx}. (Right Panel) The average momentum fraction $\langle x^n \rangle$ carried by the constituents of pion and kaon at $Q^2=16$ GeV$^2$.}
\label{jam1}
\end{figure}
\begin{figure}[ht]
\centering
\includegraphics[width=1\linewidth]{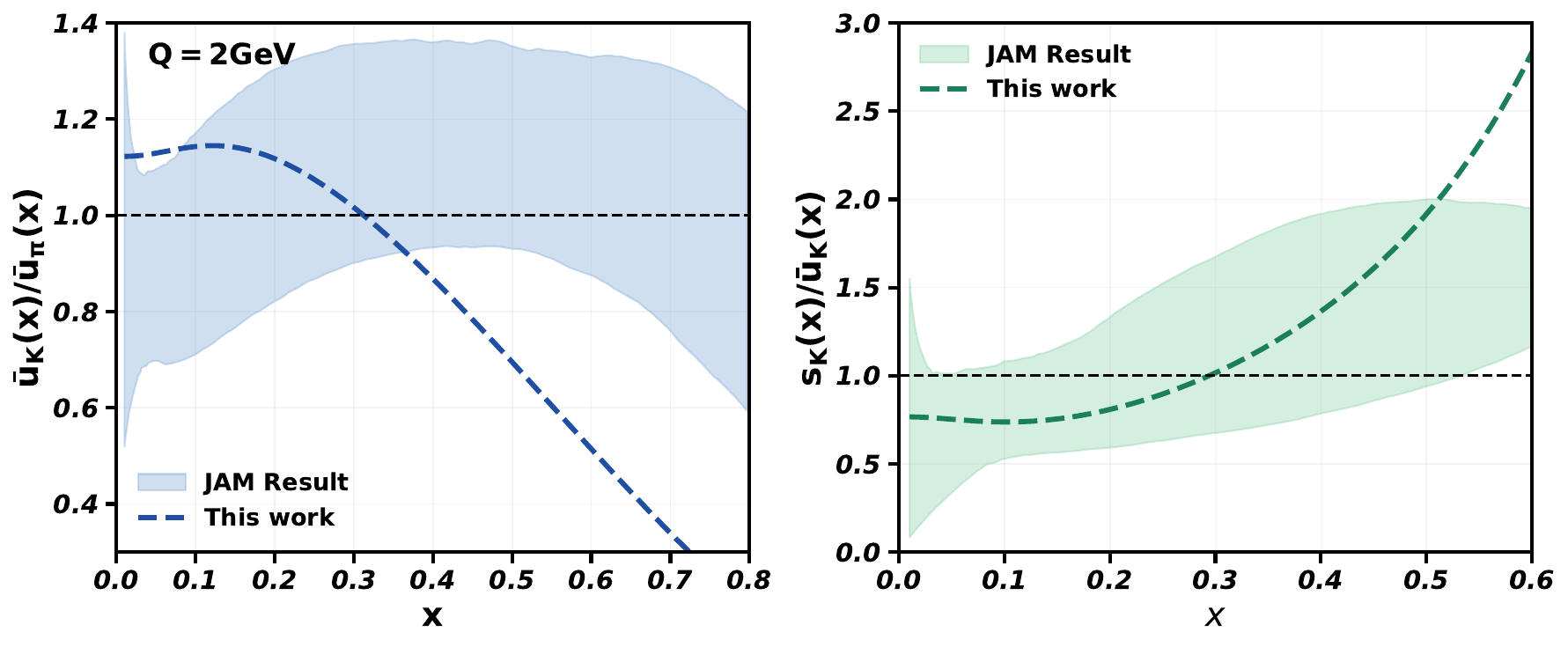}
\caption{The PDFs ratio of $\bar u_{K}(x)/\bar u_{\pi}(x)$ and $s_K(x)/\bar u_{K}(x)$ plotted in the left and right panel at $Q=2$ GeV along with the comparison with recent JAM collaboration data \cite{Barry:2025wjx}, respectively. }
\label{jam2}
\end{figure}
In Fig. \ref{jam1} (left panel), we have compared our PDFs with recent JAM collaboration data at $Q=2$ GeV \cite{Barry:2025wjx}. We observed that the constituent PDFs are found to match with up to $x=0.5$, then our PDFs gradually decrease to zero. In Fig. \ref{jam2}, we have plotted the $\bar u_{K}(x)/\bar u_{\pi}(x)$ and $s_K(x)/\bar u_{K}(x)$ ratio with recent JAM results \cite{Barry:2025wjx}. We observed that $\bar u_{K}(x)/\bar u_{\pi}(x)$ ratio is found to decrease with $x$, while $s_K(x)/\bar u_{K}(x)$ is found to increase with $x$. In Fig. \ref{jam1} (right panel), we have presented the average longitudinal momentum fraction carried by the quark antiquark of pion and kaon at $Q^2=16$ GeV$^2$. The average longitudinal momentum fraction $\langle x^n \rangle$ is calculated by
\begin{eqnarray}
    \langle x^n \rangle_{Q^2}=\int dx x^n f(x,Q^2).
\end{eqnarray}
We observed that the heavy quarks carry a longitudinal momentum fraction that is higher than that of the light quarks. At $Q=2$ GeV, $\langle x_{u(\pi)} \rangle$, $\langle x_{u(K)} \rangle$ and $\langle x_{s(K)} \rangle$ are found to be $0.23$, $0.19$ and $0.26$, respectively. At $Q^2=10$ GeV$^2$, $\langle x_{u(\pi)} \rangle$ is found to be $0.22$ compared to $0.28$ of Ref. \cite{Bourrely:2022mjf}.

%----------------------------------------------------------------
\section{Summary and Conclusion} \label{sec:sum}
%----------------------------------------------------------------
In this work, we have used the power law wave functions to study the internal structure of the pion and the kaon. First, we have calculated the distribution amplitudes of pion and kaon along with their decay constant. The decay constants of pions and kaons are found to be $114$ and $144$ MeV. We have also used the leading order ERBL evolutions to evolve the distribution amplitudes. To understand the three-dimensional structure, we have calculated the GPDs and TMDs from the quark-quark correlation functions. For the pion, both GPD and TMDs followed a symmetric behavior; however, such behavior was not seen in the case of the kaon. From GPDs, we have calculated the charge form factors and charge radii. The charge form factors are found to be in good agreement with the experimental results. The charge radii of pion and kaon are found to be $0.668$ fm and $0.704$ fm, respectively. We have also calculated the quark PDFs from the unpolarized quark TMDs by integrating over the transverse momenta. The unpolarized quark PDFs then evolved to higher energy scales using the next-to-leading order DGLAP evolution equations. The evolved PDFs have also been compared with available experimental and lattice simulation results. We have noticed that heavy quarks carry a higher longitudinal momentum fraction than light quarks. We also noticed that $60\%$ of the longitudinal momentum fraction is carried by the gluons for both pion and kaon at $Q^2=16$ GeV$^2$.

%----------------------------------------------------------------
\section*{Acknowledgments}
%----------------------------------------------------------------
S.P. acknowledges the hospitality of the International Centre for Theoretical Sciences (ICTS–TIFR), Bengaluru, where part of this work was carried out during the School for Advanced Topics in Particle Physics (SATPP 2026). H.D. would like to thank the Science and Engineering Research Board, Anusandhan-National Research Foundation, Government of India, under the scheme SERB-POWER Fellowship (Ref No. SPF/2023/000116) for
financial support.

%------------------------------------------------------------------
\section*{ORCID}
%------------------------------------------------------------------
\noindent Satyajit Puhan \orcid{0009-0004-9766-5005}
\url{https://orcid.org/0009-0004-9766-5005}

\noindent Narinder Kumar \orcid{0000-0002-5481-1162} \url{https://orcid.org/0000-0002-5481-1162}

\noindent Harleen Dahiya \orcid{0000-0002-3288-2250} \url{https://orcid.org/0000-0002-3288-2250}

%----------------------------------------------------------------


\begin{thebibliography}{0}
%----------------------------------------------------------------
%\cite{Pich:1995ua}
\bibitem{Pich:1995ua}
A.~Pich,
%``Quantum chromodynamics,''
[arXiv:hep-ph/9505231 [hep-ph]].
%70 citations counted in INSPIRE as of 22 Jan 2026

\bibitem{Puhan:2025kzz}
S.~Puhan, S.~Sharma, N.~Kumar, and H.~Dahiya,
\emph{Prog.\ Theor.\ Exp.\ Phys.} \textbf{2025}, 083B02 (2025).

%\cite{deTeramond:2018ecg}
\bibitem{deTeramond:2018ecg}
G.~F.~de Teramond \textit{et al.} [HLFHS],
%``Universality of Generalized Parton Distributions in Light-Front Holographic QCD,''
Phys. Rev. Lett. \textbf{120}, 182001 (2018).

%\cite{Meissner:2008ay}
\bibitem{Meissner:2008ay}
S.~Meissner, A.~Metz, M.~Schlegel and K.~Goeke,
%``Generalized parton correlation functions for a spin-0 hadron,''
JHEP \textbf{08}, 038 (2008).

\bibitem{Aguilar:2019teb}
A.~C.~Aguilar, \textit{et al.}
%``Pion and Kaon Structure at the Electron-Ion Collider,''
Eur. Phys. J. A \textbf{55}, 190 (2019).

%\cite{Schlumpf:1994bc}
\bibitem{Schlumpf:1994bc}
F.~Schlumpf,
%``Charge form-factors of pseudoscalar mesons,''
Phys. Rev. D \textbf{50}, 6895 (1994).

%\cite{Choi:1997qh}
\bibitem{Choi:1997qh}
H.~M.~Choi and C.~R.~Ji,
%``Relations among the light cone quark models with the invariant meson mass scheme and the model prediction of eta - eta-prime mixing angle,''
Phys. Rev. D \textbf{56}, 6010 (1997).


%\cite{Melikhov:1995xz}
\bibitem{Melikhov:1995xz}
D.~Melikhov,
%``Form-factors of meson decays in the relativistic constituent quark model,''
Phys. Rev. D \textbf{53}, 2460 (1996).

%\cite{Lepage:1980fj}
\bibitem{Lepage:1980fj}
G.~P.~Lepage and S.~J.~Brodsky,
%``Exclusive Processes in Perturbative Quantum Chromodynamics,''
Phys. Rev. D \textbf{22} (1980), 2157
%doi:10.1103/PhysRevD.22.2157
%4261 citations counted in INSPIRE as of 11 Mar 2026

%\cite{Efremov:1979qk}
\bibitem{Efremov:1979qk}
A.~V.~Efremov and A.~V.~Radyushkin,
%``Factorization and Asymptotical Behavior of Pion Form-Factor in QCD,''
Phys. Lett. B \textbf{94} (1980), 245-250
%doi:10.1016/0370-2693(80)90869-2
%1386 citations counted in INSPIRE as of 11 Mar 2026

%\cite{Chernyak:1983ej}
\bibitem{Chernyak:1983ej}
V.~L.~Chernyak and A.~R.~Zhitnitsky,
%``Asymptotic Behavior of Exclusive Processes in QCD,''
Phys. Rept. \textbf{112} (1984), 173
%doi:10.1016/0370-1573(84)90126-1
%1502 citations counted in INSPIRE as of 11 Mar 2026

%\cite{Brodsky:1973kr}
\bibitem{Brodsky:1973kr}
S.~J.~Brodsky and G.~R.~Farrar,
%``Scaling Laws at Large Transverse Momentum,''
Phys. Rev. Lett. \textbf{31} (1973), 1153-1156
%doi:10.1103/PhysRevLett.31.1153
%2036 citations counted in INSPIRE as of 13 Mar 2026

%\cite{Puhan:2024xdq}
\bibitem{Puhan:2024xdq}
S.~Puhan, N.~Kaur, A.~Kumar, S.~Dutt and H.~Dahiya,
%``Pion valence quark distributions in asymmetric nuclear matter at finite temperature,''
Phys. Rev. D \textbf{110}, 054042 (2024).


%\cite{Miyama:1995bd}
\bibitem{Miyama:1995bd}
M.~Miyama and S.~Kumano,
%``Numerical solution of Q**2 evolution equations in a brute force method,''
Comput. Phys. Commun. \textbf{94}, 185 (1996).

%\cite{Arrington:2021biu}
\bibitem{Arrington:2021biu}
J.~Arrington, \textit{et al.}
%``Revealing the structure of light pseudoscalar mesons at the electron{\textendash}ion collider,''
J. Phys. G \textbf{48}, 075106 (2021).

%\cite{Anderle:2021wcy}
\bibitem{Anderle:2021wcy}
D.~P.~Anderle, \textit{et al.}
%``Electron-ion collider in China,''
Front. Phys. (Beijing) \textbf{16}, 64701 (2021).

%\cite{Sawada:2016mao}
\bibitem{Sawada:2016mao}
T.~Sawada, W.~C.~Chang, S.~Kumano, J.~C.~Peng, S.~Sawada and K.~Tanaka,
%``Accessing proton generalized parton distributions and pion distribution amplitudes with the exclusive pion-induced Drell-Yan process at J-PARC,''
Phys. Rev. D \textbf{93}, 114034 (2016).

%\cite{Accardi:2023chb}
\bibitem{Accardi:2023chb}
A.~Accardi, \textit{et al.}
%``Strong interaction physics at the luminosity frontier with 22 GeV electrons at Jefferson Lab,''
Eur. Phys. J. A \textbf{60}, 173 (2024). 

%\cite{Adams:2018pwt}
\bibitem{Adams:2018pwt}
B.~Adams, \textit{et al.}
%``Letter of Intent: A New QCD facility at the M2 beam line of the CERN SPS (COMPASS++/AMBER),''
[arXiv:1808.00848 [hep-ex]].
%178 citations counted in INSPIRE as of 22 Jan 2026



\bibitem{Puhan:2023ekt}
S.~Puhan, S.~Sharma, N.~Kaur, N.~Kumar and H.~Dahiya,
%``T-even TMDs for the spin-0 pseudo-scalar mesons upto twist-4 using light-front formalism,''
JHEP \textbf{02}, 075 (2024).

\bibitem{Belyaev:1997iu}
V.~M.~Belyaev and M.~B.~Johnson,
%``Pion light cone wave functions and light front quark model,''
Phys. Lett. B \textbf{423}, 379 (1998).


%\cite{Puhan:2024jaw}
\bibitem{Puhan:2024jaw}
S.~Puhan, N.~Kaur and H.~Dahiya,
%``Transverse and spatial structure of light to heavy pseudoscalar mesons in light-cone quark model,''
Phys. Rev. D \textbf{111}, 014008 (2025).


%\cite{Li:2017mlw}
\bibitem{Li:2017mlw}
Y.~Li, P.~Maris and J.~P.~Vary,
%``Quarkonium as a relativistic bound state on the light front,''
Phys. Rev. D \textbf{96}, 016022 (2017).


%\cite{Choi:2007yu}
\bibitem{Choi:2007yu}
H.~M.~Choi and C.~R.~Ji,
%``Distribution amplitudes and decay constants for (pi, K, rho, K*) mesons in light-front quark model,''
Phys. Rev. D \textbf{75}, 034019 (2007).

%\cite{ParticleDataGroup:2022pth}
\bibitem{ParticleDataGroup:2022pth}
R.~L.~Workman \textit{et al.} [Particle Data Group],
%``Review of Particle Physics,''
PTEP \textbf{2022}, 083C01 (2022).

%\cite{Diehl:2003ny}
\bibitem{Diehl:2003ny}
M.~Diehl,
%``Generalized parton distributions,''
Phys. Rept. \textbf{388}, 41 (2003).

%\cite{Puhan:2025pfs}
\bibitem{Puhan:2025pfs}
S.~Puhan and H.~Dahiya,
%``Scalar, vector, and tensor form factors of pion and kaon,''
Phys. Rev. D \textbf{111}, 114039 (2025).

%\cite{NA7:1986vav}
\bibitem{NA7:1986vav}
S.~R.~Amendolia \textit{et al.} [NA7],
%``A Measurement of the Space - Like Pion Electromagnetic Form-Factor,''
Nucl. Phys. B \textbf{277}, 168 (1986).

%\cite{Mulders:1995dh}
\bibitem{Mulders:1995dh}
P.~J.~Mulders and R.~D.~Tangerman,
%``The Complete tree level result up to order 1/Q for polarized deep inelastic leptoproduction,''
Nucl. Phys. B \textbf{461} (1996), 197-237
[erratum: Nucl. Phys. B \textbf{484} (1997), 538-540]
%doi:10.1016/0550-3213(95)00632-X
%[arXiv:hep-ph/9510301 [hep-ph]].
%1068 citations counted in INSPIRE as of 27 Jan 2026

%\cite{Miyama:1995bd}
\bibitem{Miyama:1995bd}
M.~Miyama and S.~Kumano,
%``Numerical solution of Q**2 evolution equations in a brute force method,''
Comput. Phys. Commun. \textbf{94} (1996), 185-215
%oi:10.1016/0010-4655(96)00013-6
%[arXiv:hep-ph/9508246 [hep-ph]].
%147 citations counted in INSPIRE as of 23 Jan 2026

%\cite{Dally:1982zk}
\bibitem{Dally:1982zk}
E.~B.~Dally, \textit{et al.}
%``Elastic Scattering Measurement of the Negative Pion Radius,''
Phys. Rev. Lett. \textbf{48}, 375 (1982).


%\cite{JeffersonLabFpi:2000nlc}
\bibitem{JeffersonLabFpi:2000nlc}
J.~Volmer \textit{et al.} [Jefferson Lab F(pi)],
%``Measurement of the Charged Pion Electromagnetic Form-Factor,''
Phys. Rev. Lett. \textbf{86}, 1713 (2001).


%\cite{JeffersonLabFpi-2:2006ysh}
\bibitem{JeffersonLabFpi-2:2006ysh}
T.~Horn \textit{et al.} [Jefferson Lab F(pi)-2],
%``Determination of the Charged Pion Form Factor at Q**2 = 1.60 and 2.45-(GeV/c)**2,''
Phys. Rev. Lett. \textbf{97}, 192001 (2006).


%\cite{JeffersonLabFpi:2007vir}
\bibitem{JeffersonLabFpi:2007vir}
V.~Tadevosyan \textit{et al.} [Jefferson Lab F(pi)],
%``Determination of the pion charge form-factor for Q**2 = 0.60-GeV**2 - 1.60-GeV**2,''
Phys. Rev. C \textbf{75}, 055205 (2007).

%\cite{Dally:1980dj}
\bibitem{Dally:1980dj}
E.~B.~Dally,  \textit{et al.}
%``DIRECT MEASUREMENT OF THE NEGATIVE KAON FORM-FACTOR,''
Phys. Rev. Lett. \textbf{45}, 232 (1980).

%\cite{Amendolia:1986ui}
\bibitem{Amendolia:1986ui}
S.~R.~Amendolia, \textit{et al.}
%``A Measurement of the Kaon Charge Radius,''
Phys. Lett. B \textbf{178}, 435 (1986).


%\cite{Aicher:2010cb}
\bibitem{Aicher:2010cb}
M.~Aicher, A.~Schafer and W.~Vogelsang,
%``Soft-gluon resummation and the valence parton distribution function of the pion,''
Phys. Rev. Lett. \textbf{105}, 252003 (2010).

%\cite{E615:1989bda}
\bibitem{E615:1989bda}
J.~S.~Conway \textit{et al.} [E615],
%``Experimental Study of Muon Pairs Produced by 252-GeV Pions on Tungsten,''
Phys. Rev. D \textbf{39}, 92 (1989).

%\cite{Barry:2025wjx}
\bibitem{Barry:2025wjx}
P.~C.~Barry \textit{et al.} [JAM],
%``First simultaneous global QCD analysis of kaon and pion parton distributions with lattice QCD constraints,''
[arXiv:2510.11979 [hep-ph]].
%8 citations counted in INSPIRE as of 22 Jan 2026

%\cite{Bourrely:2022mjf}
\bibitem{Bourrely:2022mjf}
C.~Bourrely, W.~C.~Chang and J.~C.~Peng,
%``Pion Partonic Distributions in the Statistical Model from Pion-induced Drell-Yan and $J/\Psi$ Production Data,''
Phys. Rev. D \textbf{105}, 076018 (2022).




\end{thebibliography}
\end{document}